\documentclass[aps,12pt,tightenlines,amsmath,amssymb]{revtex4}
\usepackage{graphicx}
\usepackage{dcolumn}
\usepackage{bm}
\usepackage{float}

\begin{document}

\title{Photons to axion-like particles conversion in Active Galactic Nuclei}

\author{Fabrizio Tavecchio}
\affiliation{INAF -- Osservatorio Astronomico di Brera, Via E. Bianchi 46, I--23807 Merate, Italy}
\email{fabrizio.tavecchio@brera.inaf.it}

\author{Marco Roncadelli}
\affiliation{INFN, Sezione di Pavia, Via A. Bassi 6, I -- 27100 Pavia, Italy}
\email{marco.roncadelli@pv.infn.it}

\author{Giorgio Galanti}
\affiliation{Dipartimento di Fisica, Universit\`a dell'Insubria, Via Valleggio 11, I -- 22100 Como, Italy}
\email{gam.galanti@gmail.com}

\begin{abstract}
The idea that photons can convert to axion-like particles (ALPs) $\gamma \to a$ in or around an AGN and reconvert back to photons $a \to \gamma$ in the Milky Way magnetic field has been put forward in 2008 and has recently attracted growing interest. Yet, so far nobody has estimated the conversion probability $\gamma \to a$ as carefully as allowed by present-day knowledge. Our aim is to fill this gap. We first remark that AGN which can be detected above 100 GeV are {\it blazars}, namely AGN with jets, with one of them pointing towards us. Moreover, blazars fall into two well defined classes: BL Lac objects (BL Lacs) and Flat Spectrum Radio Quasars (FSRQs), with drastically different properties. In this Letter we report a preliminary evaluation of the $\gamma \to a$ conversion probability inside these two classes of blazars. Our findings are surprising. Indeed, while in the case of BL Lacs the conversion probability turns out to be {\it totally unpredictable} due to the strong dependence on the  values of the somewhat uncertain position of the emission region along the jet and strength of the magnetic field therein, for FSRQs we are able to make a {\it clear-cut} prediction. Our results are of paramount importance in view of the planned very-high-energy photon detectors like the CTA, HAWK, GAMMA-400 and HISCORE. 
\end{abstract}


\maketitle


\bigskip

\bigskip

\bigskip


\noindent {\it Introduction} -- Many extensions of the Standard Model of particle physics -- and chiefly among them superstring theories -- generically predict the existence of axion-like particles (ALPs) (for a review, see~\cite{alprev1,alprev2}), which are spin-zero, neutral and extremely light bosons -- to be denoted by $a$ -- closely resembling the axion (for a review, see~\cite{axion}) apart from two facts that makes them as much as model-independent as possible.

\begin{itemize} 

\item Possible couplings to fermions and gluons are discarded and only their two-photon coupling $a \gamma \gamma$ is taken into account. 

\item The ALP mass $m$ is totally unrelated to their $a \gamma \gamma$ coupling constant $1/M$. The most robust lower bound on $M$ is set by the CAST experiment CERN which yields $M > 1.14 \cdot 10^{10} \, {\rm GeV}$ for $m < 0.02 \, {\rm eV}$~\cite{cast}. Somewhat weaker bounds are $M > 1.5 \cdot 10^{10} \, {\rm GeV}$ for $m < 1 \, {\rm KeV}$ from the analysis of evolution of globular clusters~\cite{ayala} and $M > 1.9 \cdot 10^{11} \, {\rm GeV}$ for $m < 4.4 \cdot 10^{- 10} \, {\rm eV}$ from the lack of ALP detection supposedly emitted by supernova 1987A~\cite{smm}.

\end{itemize}
As a consequence, ALPs are described by the Lagrangian 
\begin{equation}
\label{t1}
{\cal L}^0_{\rm ALP} = \frac{1}{2} \, \partial^{\mu} a \, \partial_{\mu} a - \, \frac{1}{2} \, m^2 \, a^2 + \frac{1}{M} \, {\bf E} \cdot {\bf B} \, a~,
\end{equation}
where ${\bf E}$ and ${\bf B}$ denote the electric and magnetic components of the field strength $F^{\mu \nu}$. Observe that because ${\bf E}$ is perpendicular to the $\gamma$ momentum, the structure of the last term in Eq. (\ref{t1}) implies that only the component ${\bf B}_T$ transverse to the $\gamma$ momentum couples to $a$. Throughout this Letter, ${\bf E}$ is the electric field of a propagating photon while ${\bf B}$ is an {\it external} magnetic field. Accordingly, the mass matrix of the $a \gamma$ system is off-diagonal, thereby implying that the propagation eigenstates differ from the interaction eigenstates. Therefore $\gamma \leftrightarrow a$ oscillations take place much in the same way that occurs for massive neutrinos of different flavor, apart from the need of ${\bf B}$ in order to compensate for the spin mismatch~\cite{srs1,srs2}. However, in the situations to be addressed below also the one-loop QED vacuum polarization in the presence of ${\bf B}$ has to be taken into account and is described by the effective Lagrangian~\cite{HEW1,HEW2,HEW3}  
\begin{equation}
\label{t1q}
{\cal L}_{\rm HEW} = \frac{2 \alpha^2}{45 m_e^4} \, \left[ \bigl({\bf E}^2 - {\bf B}^2 \bigr)^2 + 7 \bigl({\bf E} \cdot {\bf B} \bigr)^2 \right]~,
\end{equation}
where $\alpha$ is the fine-structure constant and $m_e$ is the electron mass. So, throughout this Letter the considered ALP Lagrangian is ${\cal L}_{\rm ALP} = {\cal L}^0_{\rm ALP} + {\cal L}_{\rm HEW}$. 

In order to avoid any misunderstanding, the symbol $E$ denotes henceforth the {\it energy} rather than the electric field.

Let us now turn our attention to very-high-energy (VHE) astrophysics, namely to observed photons with energies in the range 
$100 \, {\rm GeV} < E < 100 \, {\rm TeV}$ and to their extragalactic sources, the majority of which are Active Galactic Nuclei (AGN). Generally speaking, AGN are powered by a supermassive black hole (SMBH) with $M_{\rm SMBH} \sim 10^8 - 10^9 \, M_{\odot}$ lying at the centre of a bright galaxy and accreting matter from the surrounding, which -- before disappearing into the SMBH -- heats up emitting an enormous amount of radiation. Nearly $10 \, \%$ of AGN supports two opposite relativistic jets (with a bulk Lorentz factor $\gamma \simeq 10-20$) propagating from the central regions out to distances that, in the most powerful sources, can reach $1 \, {\rm Mpc}$. Ultra-relativistic particles (leptons and/or hadrons) in the plasma carried by these jets emit non-thermal radiation extending from the radio up to the VHE band. Aberration caused by the relativistic motion makes the emission strongly anisotropic, mainly boosted in the direction of the motion. {\it Blazars} are AGN with one jet pointing -- merely by chance -- almost exactly towards the Earth. Blazars fall into two broad classes: BL Lac objects (BL Lacs) -- which represent the great majority of extragalactic sources detected in the VHE band -- and flat spectrum radio quasars (FSRQs)~\cite{urrypadovani}. As a rule, the blazar spectral energy distribution (SED) shows two broad humps, the first one peaking at low frequency -- from IR to soft-X rays, depending on the specific source -- while the second one in the $\gamma$-ray band. In BL Lacs the latter component extends to VHE, often reaching multi-TeV energies. In the widely assumed leptonic models, the VHE $\gamma$-ray emission is the result of the inverse Compton (IC) scattering of soft photons by relativistic electrons in the jet. Moreover, it is widely accepted that the dominant soft photon population derives -- through the synchrotron mechanism -- by the same electrons that scatter them into the VHE band. This is the scheme which lies at the basis of the so-called synchrotron self Compton (SSC) model~\cite{bloom,tavecchio}. 

Yet, the VHE band is plagued by the existence of the {\it extragalactic background light} (EBL) which is the light emitted by galaxies during the whole cosmic history and extends from the far-infrared to the near-ultraviolet (for a review, see~\cite{ebl}). What happens is that when a VHE $\gamma$ emitted by a distant blazar scatters off an EBL $\gamma$ it has a good chance to disappear into an $e^+ e^-$ pair~\cite{gf1,gf2}. Indeed, according to conventional physics this effect becomes dramatic even for $E$ above a few TeV (see the Fig. 1 of~\cite{dgrc}), a fact that drastically reduces the $\gamma$-ray horizon at increasing $E$. 

A breakthrough came in 2007 when it was first realized~\cite{drm} (see also~\cite{dgrx}) that $\gamma \to a \to \gamma$ oscillations taking place in intergalactic space can greatly decrease the EBL dimming for sufficiently far-away blazars and high enough $E$ provided that a large-scale magnetic field in the $0.1 - 1 \, {\rm nG}$ range exists with a domain-like structure, which is consistent with all presently available upper bounds (for a review, see~\cite{Bbounds}). Why this happens can be understood in an intuitive fashion (discarding cosmological effects for simplicity). Photon-ALP oscillations give a photon a split personality: as it propagates from the blazar to us, it behaves sometimes as a {\it true} photon and sometimes as an ALP. When it propagates as a photon it undergoes EBL absorption, but when it propagates as an ALP it does {\it not}. Therefore, the effective photon mean free path in extragalactic space $\lambda_{\gamma, {\rm eff}} (E)$ is {\it larger} than $\lambda_{\gamma} (E)$ as predicted by conventional physics. Correspondingly, the photon survival probability becomes $P_{\gamma \to \gamma} (E) = {\rm exp} \bigl(- D_s/\lambda_{\gamma, {\rm eff}} (E) \bigr)$, where $D_s$ is the blazar distance. So, because of the exponential dependence on the mean free path even a small increase of $\lambda_{\gamma, {\rm eff}} (E)$ with respect to ${\lambda}_{\gamma}(E)$ produces a large enhancement of $P_{\gamma \to \gamma} (E)$, thereby giving rise to a drastic reduction of the EBL dimming. 

Before proceeding further, a remark is in order. From time to time a tension between the predicted EBL level causing photon absorption and observations in the VHE range has been claimed~\cite{2000protheroe,aharonian2006}, but a subsequent better determination of the EBL properties has shown that no problem exists. Actually, after a long period of uncertainty on the EBL precise properties, nowadays a convergence seems to be reached~\cite{ebl}, well represented e.g. by the models of Franceschini, Rodighiero and Vaccari (FRV)~\cite{frv} and of  Dom\'inguez {\it et al.}~\cite{dominguez}. Nevertheless, it has been claimed that VHE observations require an EBL level even lower than that predicted by the minimal EBL model normalized to the galaxy counts only~\cite{kneiske}. This is the so-called {\it pair-production anomaly}, which is based on the Kolmogorov test and so does not rely upon the estimated errors. 
It has thoroughly been quantified by a global statistical analysis of a large sample of observed blazars, showing that measurements in the regime of large optical depth deviate by 4.2 $\sigma$ from measurements in the optically thin regime~\cite{hornsmeyer2012}. Systematic effects have been shown to be insufficient to account for such the pair-production anomaly, which looks therefore real. Actually, the discovery of new blazars at large redshift like the observation of PKS 1424+240 have strengthened the case for the pair-production anomaly~\cite{hmpks}. Quite recently, the existence of the pair-production anomaly has been questioned by using a new EBL model and a $\chi^2$ test, in which errors play instead an important role~\cite{biteau}. Because the Kolmogorov test looks more robust in that it avoids taking errors into account, we tend to believe that the pair-production anomaly is indeed at the level of 4.2 $\sigma$. It looks tantalizing that for a suitable choice of the free parameters it has been shown that the mechanism discussed above provides a solution to the pair-production anomaly~\cite{hornsmeyer2012,ppanomaly1}. An even more amazing fact is that for the {\it same} choice of the free parameter also the observed redshift-dependence of the blazar spectra is naturally explained~\cite{GRDB2015}.

Coming back to our main line of development, as a follow-up of the previous proposals that $\gamma \to a$ conversions occur in AGN~\cite{bischeri,hochmuth}, a complementary scenario was put forward in 2008~\cite{she}. Schematically, VHE photons are simply {\it assumed} to substantially or even maximally convert to ALPs inside a blazar, so that the emitted flux can consist in up to $1/3$ of ALPs and in $2/3$ of photons. ALPs travel unaffected by the EBL and when they reach the Milky Way (MW) can convert back to photons in the MW magnetic field. Clearly, the amount of back-conversion strongly depends on the galactic coordinates of the blazar, since the morphology of the MW magnetic field is quite complicated and by no means isotropic. Evidently also in this case the EBL dimming is drastically reduced. Basically the same idea has been taken up subsequently~\cite{ph1,ph2}. Unfortunately, either the blazar has not been modelled at all~\cite{bischeri,she} or an incorrect domain-like structure model for the jet magnetic field is assumed~\cite{ph1,ph2}. 

Prompted by the appearance of very recent papers addressing the considered scenario in connection with the upcoming Cherenkov Telescope Array (CTA)~\cite{bruntro1,bruntro2,mmc,mc}, we have decided to report a preliminary evaluation of the $\gamma \to a$ conversion probability $P_{\gamma \to a} (E)$ inside the two classes of blazars as carefully as possible consistently with the presently available knowledge. So, the aim of this Letter is basically to speed up the presentation of our results. A more thorough analysis along with all relevant calculations will be the subject of a future much more detailed paper.

\

\noindent {\it BL Lacs} -- They are the simplest blazars, and so we better start from them. Denoting by $y$ the coordinate along the jet axis, in order to achieve our goal two quantities are needed where the photon/ALP beam propagates: the transverse magnetic field ${\bf B}_T ( y )$ and the electron density $n_e ( y )$ profiles. Because the electrons are accelerated in shocks generated in the flow, the VHE photons are produced in a well-localized region ${\cal R}_{{\rm VHE}}$ pretty far from the central engine. So, four crucial parameters are: the distance $d_{{\rm VHE}}$ of ${\cal R}_{{\rm VHE}}$ from the centre, the size of ${\cal R}_{{\rm VHE}}$, and the values of $B_{T, {\rm VHE}}$ and $n_{e, {{\rm VHE}}}$ inside it. The SSC diagnostics applied to the SED of BL Lacs~\cite{tavecchio2010} provides the main physical quantities concerning ${\cal R}_{{\rm VHE}}$. They are $B_{T, {\rm VHE}} = 0.1 - 1 \, {\rm G}$ and $n_{e, {{\rm VHE}}} \simeq 5 \cdot 10^{4} \, {\rm cm}^{-3}$, leading to a plasma frequency of $8.25 \cdot 10^{- 9} \, {\rm eV}$. The quantity $d_{{\rm VHE}}$ is difficult to determine directly, because the current instrumental spatial resolution is still too poor. A common indirect way consists in inferring $d_{{\rm VHE}}$ from the size $R_{{\rm VHE}}$ of ${\cal R}_{{\rm VHE}}$, assumed to be a measure of the jet cross-section, derived in turn from spectral models and the observed variability timescale. Typical values lie in the range $10^{15} - 10^{16} \, {\rm cm}$. Whenever the jet aperture angle $\theta _{\rm jet}$ is measurable -- which is certainly the case for BL Lacs at a relatively large distance -- it is generally found $\theta _{\rm jet}  \simeq 0.1$ rad, so that under the assumption of a simple conical geometry for the jet it follows that $d_{{\rm VHE}} = R_{{\rm VHE}} / \theta_{\rm jet} \simeq 10^{16} -10^{17} \, {\rm cm}$. Beyond ${\cal R}_{{\rm VHE}}$ photons travel outwards unimpeded until they leave the jet with a typical length of $1 \, {\rm kpc}$ and propagate into the host galaxy. Given the fact that $d_{{\rm VHE}}$ is a fairly large quantity, the component of ${\bf B}$ relevant for us is the toroidal part which is transverse to the jet axis and goes like $y^{- 1}$~\cite{bbr1984}. The same conclusion follows from the conservation of the magnetic luminosity if the jet conserves its speed~\cite{ghisellini2009}. Moreover, recent work has succeeded to observationally characterize the ${\bf B}$ structure over distances in the range $0.1 - 100 \, {\rm pc}$ in several jets of BL Lacs  through polarimetric studies, showing unambiguously that in BL Lacs ${\bf B}$ is indeed substantially {\it ordered} and predominantly {\it traverse} to the jet~\cite{pudritz2011}. We stress that in particular these results are inconsistent with a domain-like structure of ${\bf B}$ in the jet as assumed e.g. in~\cite{ph1,ph2}. Turning next to the electron density, under the usual assumption that the jet has a conical shape we expect that it goes like $y^{- 2}$. Whence 
\begin{equation}
\label{Bjet}
B_T ( y ) = \frac{B_{T,{{\rm VHE}}} \ d_{{\rm VHE}}}{y}~, \ \ \ \ n_e ( y ) = \frac{n_{{\rm VHE}} \ d^2_{{\rm VHE}}}{y^2}~,
\end{equation}
for $y > d_{{\rm VHE}}$. Observe that Eqs. (\ref{Bjet}) holds true in a frame co-moving with the jet, so that the transformation to a fixed frame is effected by $E \to \gamma E$. We remark that this relation is strictly true if  the jet is observed at an angle $\theta_v=1/\gamma$ with respect to the jet axis. More generally, the transformation reads $E\to E\,\delta$, where $\delta$ is the {\it relativistic Doppler factor} (for details, see~\cite{urrypadovani}). Here we have $\gamma = 15$.

\

\begin{figure}[H]    
\centering
\includegraphics[width=1.20\textwidth,height=1.30\textwidth]{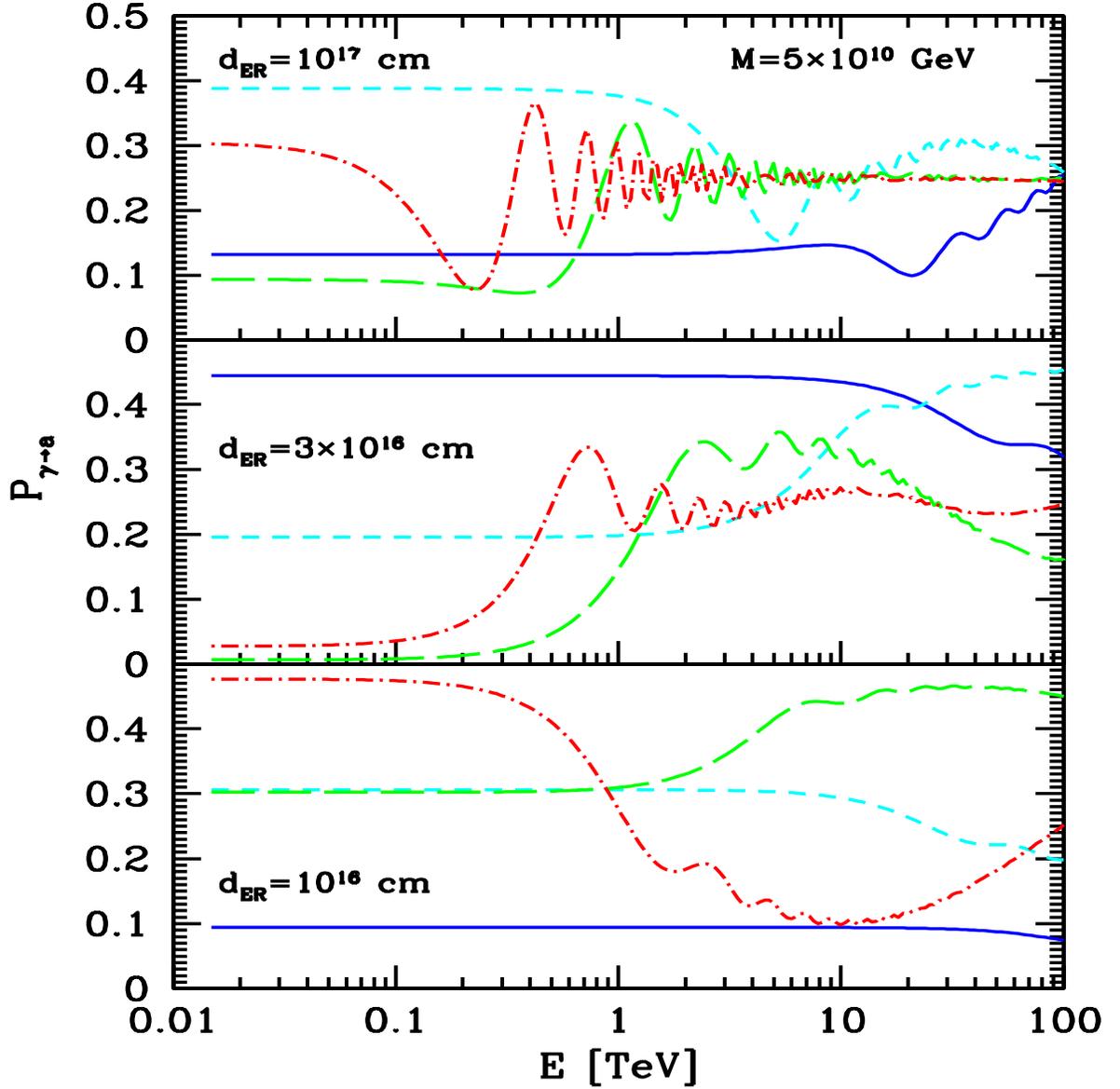}
\caption{\label{fig:nohost_1kpc_M05} Plot of $P_{\gamma \to a} (E)$ for a BL Lac taking $M = 5 \cdot 10^{10} \, {\rm GeV}$. The different curves correspond to $B=0.1 \, {\rm G}$ (solid blue), $0.2 \, {\rm G}$ (dashed cyan), $0.5 \, {\rm G}$ (long dashed, green) and $1 \, {\rm G}$ (dot-dashed, red). The three panels correspond to three values of the distance of the emitting region, namely $d_{\rm VHE} =10^{16} \, {\rm cm}$ (bottom), $3 \cdot 10^{16} \, {\rm cm}$ (middle), $10^{17} \, {\rm cm}$ (upper).}
\end{figure}  

\begin{figure}[H]    
\centering
\includegraphics[width=1.20\textwidth,height=1.30\textwidth]{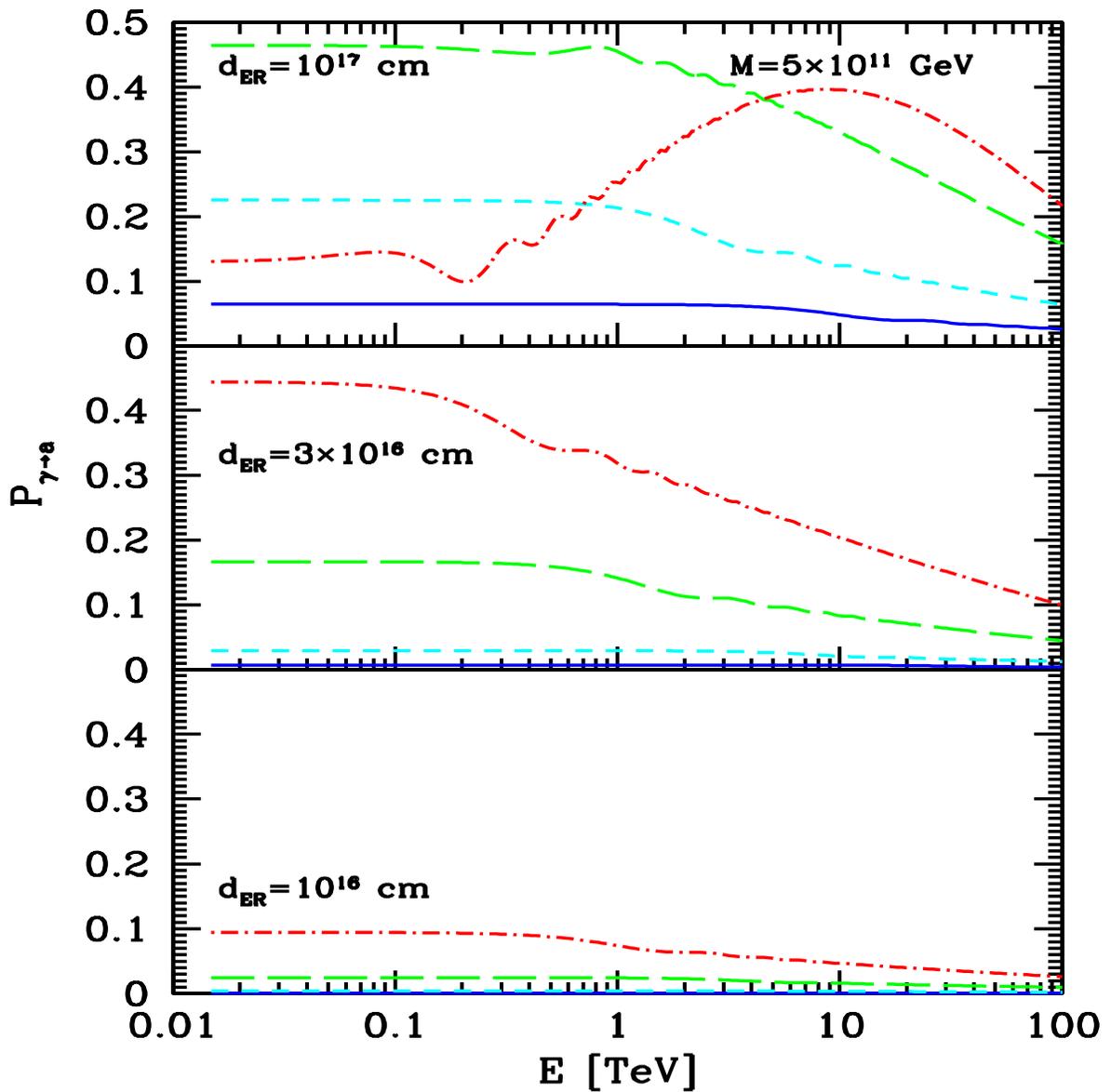}
\caption{\label{fig:nohost_1kpc_M5} Plot of $P_{\gamma \to a} (E)$ for a BL Lac taking $M = 5 \cdot 10^{11} \, {\rm GeV}$. The different curves correspond to $B=0.1 \, {\rm G}$ (solid blue), $0.2 \, {\rm G}$ (dashed cyan), $0.5 \, {\rm G}$ (long dashed, green) and $1 \, {\rm G}$ (dot-dashed, red). The three panels correspond to three values of the distance of the emitting region, namely $d_{\rm VHE} =10^{16} \, {\rm cm}$ (bottom), $3 \cdot 10^{16} \, {\rm cm}$ (middle), $10^{17} \, {\rm cm}$ (upper).}
\end{figure}  

\

\noindent {\it FSRQs} -- These are the most powerful blazars and are in a sense a more complicated version of BL Lacs. The additional components are: (1) the broad line region (BLR) consisting in a spherical shell of many clouds photo-ionized by the radiation from the matter accreting onto the SMBH, located at about $d_{{\rm BLR}} \simeq10^{18} \, {\rm cm}$ from the centre and rapidly rotating about it; (2) a dusty torus reprocessing part of the above radiation in the infrared band; (3) the radio lobes consisting in a hot non-thermal plasma inflated where the jets collide with the extragalactic gas. Because both the BLR and the dusty torus lie {\it beyond} ${\cal R}_{{\rm VHE}}$ and are quite rich of ultraviolet and infrared photons, respectively, they give rise to a huge absorption of $\gamma$ rays with $E_{\gamma} > 10 - 20 \, {\rm GeV}$ through the same $\gamma \gamma \to e^+ e^-$ process considered above. On the other hand, the lobes -- being magnetized -- represents a further conversion region. The jets in FSRQs are longer and less prone to instabilities than those of the weaker BL Lacs. Presently, concerning the VHE $\gamma$-ray emission region ${\cal R}_{\rm VHE}$ we take $d_{\rm VHE}$ larger by a factor of 3 as compared to the BL Lac case, based on the larger variability time scales~\cite{bfsrq}. The modeling of the SED with state-of-the-art emission models provides $B_{T,{\rm VHE}} = 1 - 10 \, {\rm G}$ and $n_{\rm VHE} \simeq 10^4 \, {\rm cm}^{- 3}$~\cite{bfsrq}. The geometry and the intensity of ${\bf B}$ in the jet beyond ${\cal R}_{{\rm VHE}}$ are far less clear than in the case of BL Lacs. In fact, there are indications that ${\bf B}$ has a globally ordered structure, but its inclination angle $\varphi$ with respect to the jet axis does not have a unique value for all sources, actually covering the whole interval $0 - 90^{\circ}$. For definiteness, we assume the same profiles of $B_T ( y )$ and $n_e ( y )$ as in Eq. (\ref{Bjet}), taking $\varphi = 45^{\circ}$ and $\gamma = 10$. Radio polarimetric observations yield a good amount of information about the structure and the intensity of ${\bf B}$ in the radio lobes. Specifically, one gets a turbulent ${\bf B}$ which can be modelled as a domain-like structure with homogenous strength $B = 10 \, {\mu}{\rm G}$, coherence length $10 \, {\rm kpc}$ and random orientation in each domain.  

\

\begin{figure}[H]    
\centering
\includegraphics[width=1.20\textwidth,height=1.30\textwidth]{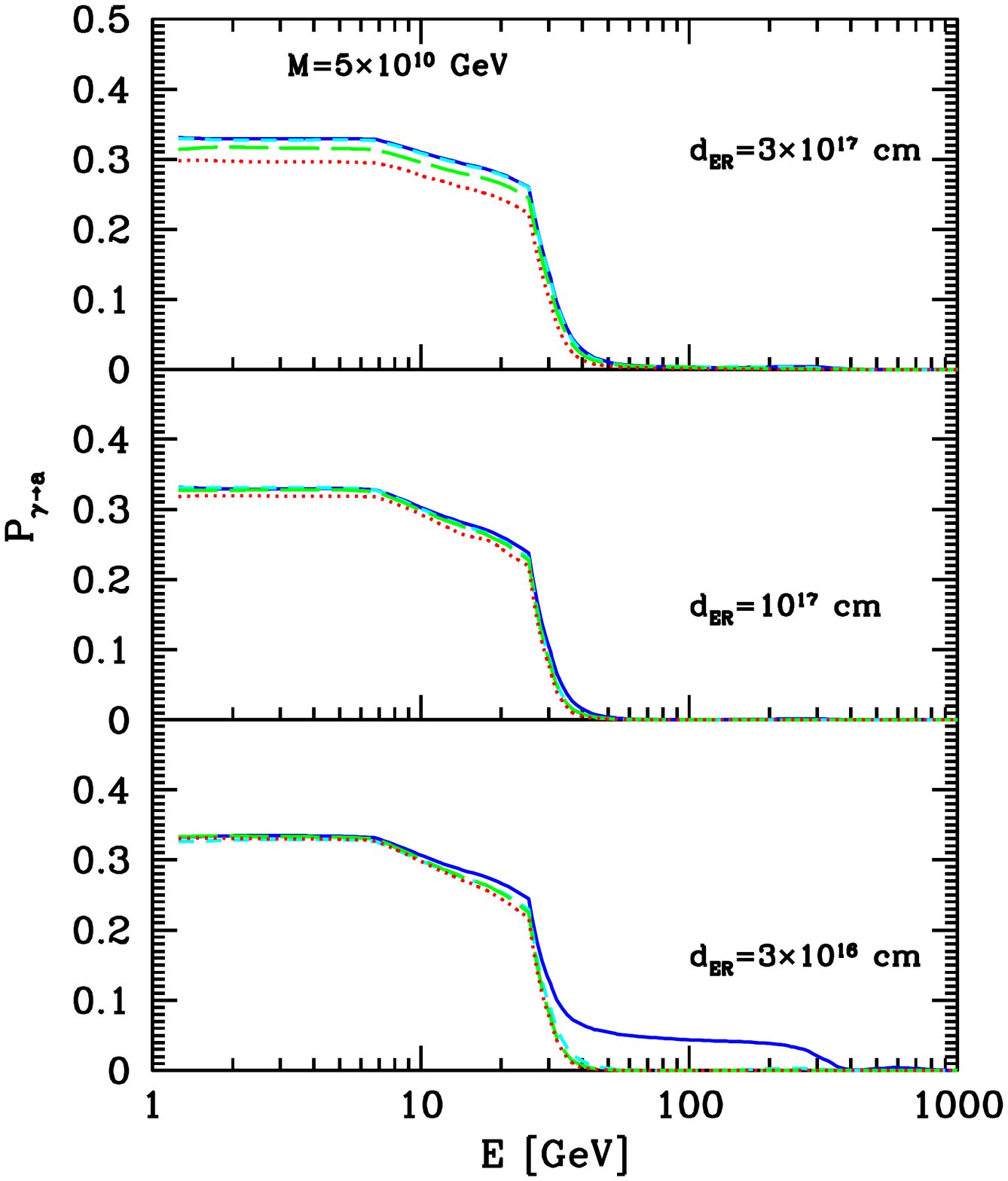}
\caption{\label{fig:pks_M05} Plot of $P_{\gamma \to a} (E)$ for a FSRQ taking $M = 5 \cdot 10^{10} \, {\rm GeV}$. The different curves correspond to $B=1 \, {\rm G}$ (solid blue), $2 \, {\rm G}$ (dashed cyan), $5 \, {\rm G}$ (long dashed, green) and $10 \, {\rm G}$ (dot red). The three panels correspond to three values of the distance of the emitting region, namely $d_{\rm VHE} =3 \cdot 10^{16} \, {\rm cm}$ (bottom), $10^{17} \, {\rm cm}$ (middle), $3 \cdot 10^{17} \, {\rm cm}$ (upper).}
\end{figure}  

\begin{figure}[H]    
\centering
\includegraphics[width=1.20\textwidth,height=1.30\textwidth]{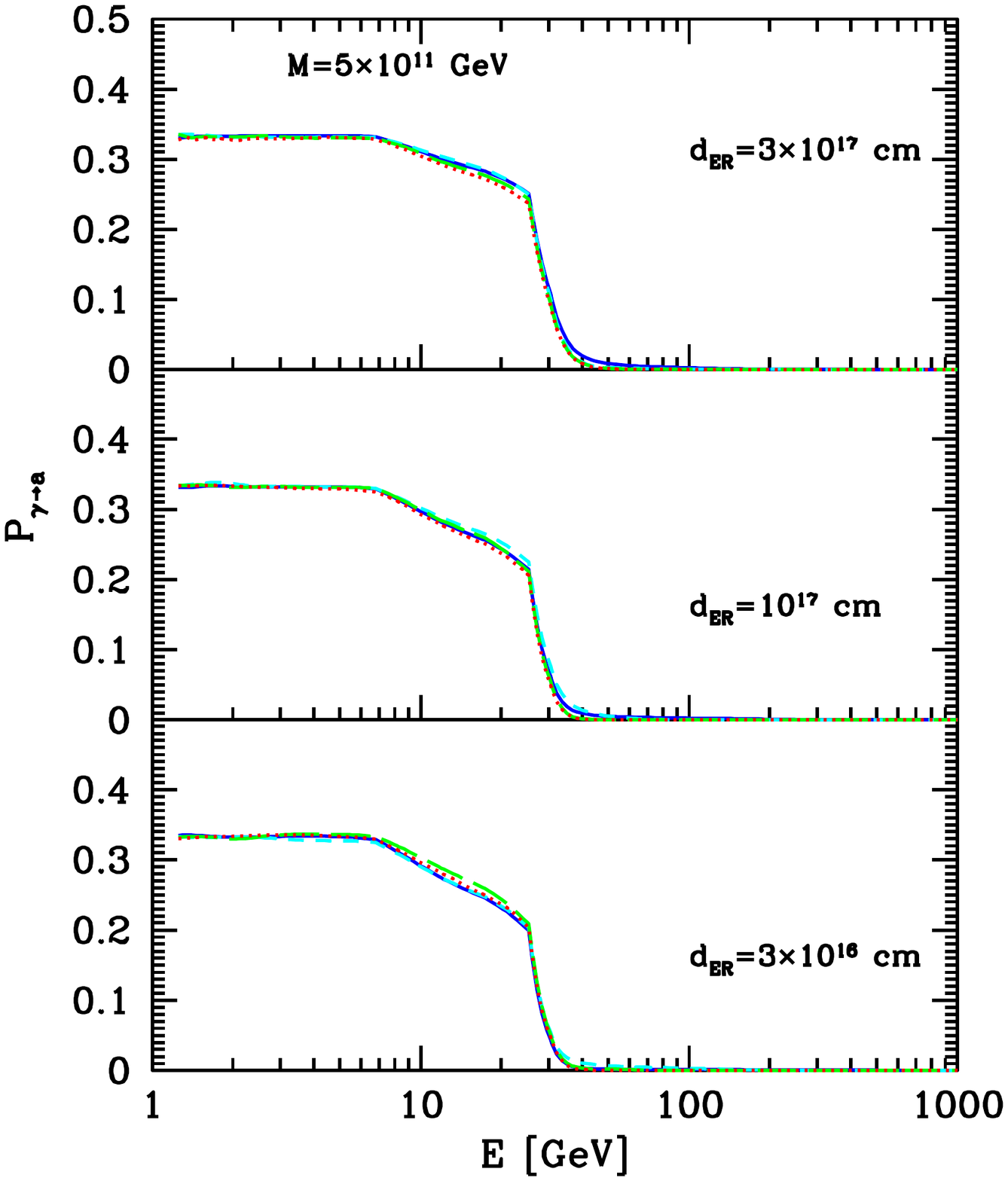}
\caption{\label{fig:pks_M5} Plot of $P_{\gamma \to a} (E)$ for a FSRQ taking $M = 5 \cdot 10^{11} \, {\rm GeV}$. The different curves correspond to $B=1 \, {\rm G}$ (solid blue), $2 \, {\rm G}$ (dashed cyan), $5 \, {\rm G}$ (long dashed, green) and $10 \, {\rm G}$ (dot red). The three panels correspond to three values of the distance of the emitting region, namely $d_{\rm VHE} =3 \cdot 10^{16} \, {\rm cm}$ (bottom), $10^{17} \, {\rm cm}$ (middle), $3 \cdot 10^{17} \, {\rm cm}$ (upper).}
\end{figure}  

\

\noindent {\it Results} -- Because of lack of space, we cannot report the explicit calculation of the $\gamma \to a$ conversion probability $P_{\gamma \to a} (E)$ which is anyway a straightforward application of the technique discussed in great detail 
in~\cite{dgrx}. We assume as benchmark values $m \le 10^{- 9} \, {\rm eV}$ as well as $M = 5 \cdot10^{10} \, {\rm GeV}$ and 
$M = 5 \cdot10^{11} \, {\rm GeV}$. Basically, our results can be summarized as follows. 

Owing to the leading role played by the QED term, in the case of BL Lacs $P_{\gamma \to a} (E)$ shows a rather complex behaviour and a strong dependence on $B_{T,{\rm VHE}}$ and $d_{\rm VHE}$, as shown in Figs.~\ref{fig:nohost_1kpc_M05} and ~\ref{fig:nohost_1kpc_M5}. As a consequence, $P_{\gamma \to a} (E)$ turns out to be {\it intrinsically unpredictable}. In addition, as $M$ decreases only for the largest considered values of the magnetic field takes the conversion probability sizable values and no oscillatory behaviour shows up.

On the contrary, for FSRQs due to the efficient $\gamma \leftrightarrow a$ oscillations in the radio lobes -- which actually leads to the equipartition among the three degrees of freedom -- the peculiar features exhibited by BL Lacs get smoothed out and below $20 \, {\rm GeV}$ we get $P_{\gamma \to a} (E) = 1/3$ regardless of the value of $M$. Above $20 \, {\rm GeV}$ instead the above-mentioned absorption leads to a drastic reduction of the emitted ALP flux. Altogether, in the case of FSRQs we make a clear-cut prediction which is exhibited in Figs.~\ref{fig:pks_M05} and~\ref{fig:pks_M5}, which is again almost independent of the value of $M$. 

Our results are of great importance for the planned very-high-energy detectors like the CTA, HAWK, GAMMA-400 and HISCORE, and for those based on the techniques discussed in~\cite{avignone1,avignone2,avignone3}. Moreover, we stress that all analyses of the scenario of $\gamma \to a$ conversion in a blazar and $a \to \gamma$ reconversion in the MW  should be properly revised according to the present conclusions.

\

\noindent {\it Acknowledgments} -- We thank Alessandro De Angelis, Luigina Feretti and Marcello Giroletti for useful discussions. F. T. acknowledges contribution from a grant PRIN-INAF-2011. The work of M. R. is supported by INFN TAsP and CTA grants.

\bibliography{mybibfile}

\end{document}